\documentclass[11pt,osajnl2,preprint,showpacs,superscriptaddress]{revtex4}  
\usepackage[draft]{hyperref} 
\newcommand{\be}{\begin{equation}}
\newcommand{\ee}{\end{equation}}

\newcommand{\ba}[1]{\left(\begin{array}{#1}}
\newcommand{\ea}{\end{array}\right}
\usepackage{times}
\usepackage{amsmath}
\usepackage{amssymb}
\usepackage{mathrsfs}
\usepackage{array}
\usepackage{latexsym}
\usepackage{setspace}
\usepackage{subfigure}
\usepackage{epsf}
\usepackage{psfrag}
\usepackage{rotating}
\usepackage{url}
\usepackage{fancybox}
\usepackage[usenames]{color}
\usepackage{rotate}
\usepackage{lscape}

\begin{document}
\title{{\bf Loss of exchange symmetry in multiqubit states under Ising chain evolution}}

\author{Sudha}
\email{arss@rediffmail.com}
\affiliation{Department of Physics, Kuvempu University, 
Shankaraghatta-577 451, India} 
\author{Divyamani B.G}
\affiliation{Department of Physics, Kuvempu University, 
Shankaraghatta-577 451, India} 
\author{ A. R. Usha Devi} 
\affiliation{Department of Physics, Bangalore University, 
Bangalore-560 056, India}
\begin{abstract}
Keeping in view the importance of exchange symmetry aspects in studies on spin squeezing of multiqubit states, we show that Ising type Hamiltonian does not retain the exchange symmetry of initially symmetric multiqubit states. Specifically, we show that all $N$-qubit states ($N\geq 5$) obeying permutation symmetry lose their symmetry after evolution with one dimensional Ising chain with nearest neighbor interactions. Among 4-qubit symmetric pure states, only W type states retain their symmetry under time evolution with Ising Hamiltonian.  
\end{abstract}
\maketitle
\section{Introduction}
Multiqubit states that are symmetric under interchange of particles form an important class among quantum states due to their experimental significance 
and mathematical elegance~\cite{sym1,sym2,sym3}. They are the quantum states obeying exchange symmetry and the 
$N$-qubit symmetric states belong to the $N+1$ dimensional subspace of the $2^N$ dimensional Hilbert space, the 
subspace being the maximal multiplicity space of the collective angular momentum. In general, multi-atom systems 
that are symmetric under permutation of the particles allow for an elegant description in terms of the 
collective variables of the system. 

Atomic spin squeezed states~\cite{kitueda1,wineland,gsa,asmolmer,ssI1,xw,sym2} are quantum correlated states with reduced 
fluctuations in one of the collective spin components and they have possible applications in atomic interferometers and
high precision atomic clocks.  
If ${\bf J}_i= \sum_{\alpha=1}^N\sigma_{\alpha i}/2$,  $i=x,\,y,\,z$ denotes the components of the collective angular momentum operator of an $N$ qubit system, one defines~\cite{kitueda1} 
the spin squeezing parameter $\xi$ as
\be
\xi^2=\frac{2(\Delta{\bf J}_{\bot})^2}{J}; \ \ J=\frac{N}{2}
\ee 
Here the subscript $\bot$ refers to an axis perpendicular to the mean spin $<\vec{\bf J}>$ where the minimal value of the variance $(\Delta{\bf J}_{\bot})^2$ is obtained. The system is said to be spin squeezed when the parameter $\xi^2$ is less than 1. 
The relationship between spin squeezing and quantum entanglement has been an interesting area of study~\cite{kitueda2,sspe2} and it has been 
shown that for a two qubit symmetric state, spin squeezing is equivalent to its bipartite entanglement~\cite{kitueda2}. An extension of this result to symmetric multiqubit systems shows that the presence of spin squeezing essentially reflects pairwise entanglement~\cite{sspe2}. Eventhough spin squeezing serves only as a sufficient condition for pairwise entanglement in arbitrary symmetric multiqubit systems, for a special class of symmetric multiqubit systems it was shown that spin squeezing is a necessary and sufficient condition for pairwise entanglement~\cite{sspe2}. A relation between the squeezing parameter $\xi$ and concurrence, a measure of two-qubit entanglement, has also been obtained~\cite{sspe2}.  

With the observation that the detection of spin squeezing forms a useful diagnostic tool in the early stages of the construction of a quantum computer~\cite{ssI1}, the spin squeezing produced in several models of interacting spins has been studied~\cite{ssI1,sspe2}. Ising type Hamiltonian with nearest neighbor interactions~\cite{ssI1} is one among the interaction models considered in these studies. The one dimensional Ising type Hamiltonian with $N$ spins and a constant coupling between any two nearest neighbors~\cite{ssI1} is given by  
\be
{\bf H}=\frac{\hbar\chi}{4} \sum_{\alpha=1}^{N}  
\sigma_{\alpha x}\sigma_{\alpha+1 x} 
\ee
where we identify the $(N+1)^{\rm th}$ spin with the first one in the chain.
Here $\sigma_{\alpha x}$ and $\sigma_{\alpha+1 x}$ are the Pauli spin matrices for the spin at site $\alpha$, 
$\alpha+1$ respectively and $\chi$ is a constant characterizing the coupling strength between any two nearest 
neighbors in the chain. It is not difficult to see that, ${\bf J}_x$=$\frac{1}{2}\sum_{\alpha=1}^{N}\sigma_{\alpha x}$, the $x$ component of the collective angular momentum operator is a constant of motion as it commutes with the Hamiltonian 
${\bf H}$ of the system.   
The Ising type Hamiltonian  arises in recent proposals for quantum computation with atoms in optical lattices~\cite{ol1,ol2}. In these proposals, the atom interacts with nearest neighbors and it has been shown that this interaction produces spin squeezing.  Spin squeezed states are routinely produced in several laboratories as they are quite experimentalist-friendly and in addition to the practical applications of spin squeezing such as atomic clocks, they provide a demonstration of the entangling capabilities of the system. Thus studies on spin squeezed states and the interaction models that produce spin squeezing form a prominent area of study. 

At this juncture, it is important to recall that spin-squeezing is a property 
defined for quantum states obeying exchange symmetry~\cite{kitueda1, kitueda2} and while examining the spin squeezing behaviour of symmetric multiqubit states interacting through a particular Hamiltonian model, one has to know whether the exchange symmetry of the state is affected by the interacting Hamiltonian or not. As any analysis of spin squeezing in non-symmetric states is bound to give invalid results, permutation symmetry of a multiqubit state has to be ascertained before examining its spin squeezing nature. The main motivation of the present work is to show that permutation symmetry of an initially symmetric multiqubit state cannot be taken for granted while considering  its time evolution with different interaction models. As exchange symmetry aspects are not given consideration in studies on spin squeezed states generated through time evolution of an initially symmetric multiqubit state with Ising type interaction models~\cite{ssI1,asmolmer,ssI2}, an exploration of this aspect in an explicit manner has to be given due importance. We carry out one such study in this article. 
In fact, we show that none of the $N$ qubit ($N\geq 4$) symmetric states, except the 4 qubit W state, retain their exchange symmetry under evolution with Ising type Hamiltonian, one of the important interaction models considered for spin squeezing studies~\cite{ssI1}.  

The paper is organized as follows: In section~II of the paper, starting with the one dimensional Ising Hamiltonian corresponding respectively to 2, 3 and 4 qubits, we explicitly show that the initially symmetric pure states retain their exchange symmetry under unitary evoltuion with Ising chain Hamiltonian in the 2 and 3 qubit case. We also show that all other four qubit symmetric states, except the states of W type, lose their permutation symmetry under Ising chain evolution. The loss of exchange symmetry in the respective cases is evident from the fact that the evolved states  
do not remain in the symmetric subspace of the $N$ qubit states. We have summarized our results in Tables~I and II.  Section~III has concluding remarks
        
\section{Exchange Symmetry of multiqubit states under Ising chain interaction}
\label{section2}

An Ising chain with $N=2$ spins is given by ${\bf 
H}=\frac{\hbar\chi}{4}(\sigma_{1x}\sigma_{2x}+\sigma_{2x}\sigma_{1x})$. The states spanning the $2+1=3$ 
dimensional symmetric subspace~\footnote{The symmetric subspace of a $N$-qubit system is the maximal multiplicity 
space of the collective angular momentum $j=N/2$. The basis states of the symmetric subspace are the 
$|j,m\rangle$ states, $j=N/2$, $m=-N/2$ to $N/2$.} are the spin-up ($|00\rangle$), spin-down ($|11\rangle$) and 
the state $\vert\psi\rangle=\frac{|01\rangle+|10\rangle}{\sqrt{2}}$. As $\sigma_{1x}$, $\sigma_{2x}$ correspond 
to spin flip operation on $1^{\rm st}$ and $2^{\rm nd}$ qubit respectively, it is easy to see that, 
\[
{\bf H}|00\rangle \propto  |11\rangle, \ \   {\bf H}|11\rangle \propto |00\rangle, \ \  {\bf H}\vert\psi\rangle 
\propto \vert\psi\rangle 
\]
Hence repeated application of $\bf H$ on these states results in the same states. Thus the action of the 
time-evolution operator ${\bf U}=\exp (-i{\bf H}t/\hbar)$ on basis states of the symmetric subspace results in 
their linear combination, ensuring the symmetry of the 2-qubit Dicke states under Ising chain evolution. 

Considering the Ising chain with $N=3$ spins, we have 
\be
{\bf H}=\frac{\hbar\chi}{4}(\sigma_{1x}\sigma_{2x}+\sigma_{2x}\sigma_{3x}+\sigma_{3x}\sigma_{1x}).
\ee
The set of all symmetric 3 qubit states is spanned by the $3+1=4$ basis states (3 qubit Dicke states)
\begin{eqnarray}
\vert\psi_1\rangle&=& |3/2, 3/2\rangle=|000\rangle \nonumber \\ 
\vert\psi_2\rangle&=& |3/2, 1/2\rangle=\frac{|001\rangle+|010\rangle+|100\rangle}{\sqrt{3}} \nonumber \\  
\vert\psi_3\rangle&=&|3/2, -1/2\rangle=\frac{|110\rangle+|101\rangle+|011\rangle}{\sqrt{3}} \\ 
\vert\psi_4\rangle&=&|3/2, -3/2\rangle=|111\rangle. \nonumber
\end{eqnarray} 
Our main task here is to check whether an arbitrary symmetric 3-qubit state retains its symmetry under Ising 
chain evolution. For this, we need to examine whether the basis states of the symmetric subspace 
of three qubits remain symmetric after interaction with 1D Ising chain. 

Though the exchange symmetry of each of the 3-qubit Dicke states after interaction with the 1D Ising chain can be checked 
by inspection as is done for the 2 qubit case, it is easier to examine whether the evolved states 
$\vert\psi'_\alpha\rangle={\bf U}\vert\psi_\alpha\rangle$ ($\alpha=1,\,2,\,3,\,4$), ${\bf U}=\exp (-i{\bf H}t/\hbar)$ being the unitary operator corresponding to 3-qubit Ising chain Hamiltonian, remain in the symmetric 
subspace.  It is a simple matter to notice that the evolved states $\vert\psi'_\alpha\rangle={\bf 
U}\vert\psi_\alpha\rangle$ ($\alpha=1,\,2,\,3,\,4$) remain in the symmetric subspace iff 
$\vert\psi'_\alpha\rangle$ is expressible as $\vert\psi'_\alpha\rangle=\sum_{m} c_{m}|\frac{3}{2}\,,\,m\rangle$ 
such that $\sum_{m}|c_m|^2=1$. In fact,   $c_m=\langle\psi'_\alpha|\frac{3}{2},m\rangle$ and we need to evaluate 
the set of coefficients $c_m=\langle\psi'_\alpha|\frac{3}{2}\,,\,m\rangle, \ 
m=-\frac{3}{2},\,-\frac{1}{2},\,\frac{1}{2},\,\frac{3}{2}$ for each $\alpha$ ($\alpha=1,\,2,\,3,\,4$). The 
coefficients $c_m$ for each of the states 
$\vert\psi_\alpha\rangle$ are given in Table~I and it is evident that the symmetry of the 3 qubit Dicke states 
is unhampered by Ising chain interaction. 

Starting with an Ising chain with four spins, we evaluate the corresponding unitary time-evolution operator and 
it is given by 
\be
\label{eq:unitary1}
{\bf U}=\exp \left( -\frac{i{\bf H}t}{\hbar} \right)={\bf I}+{\bf A}^2(\cos \chi t-1)-i{\bf A} \sin \chi t
\ee   
Here ${\bf H}=\frac{\hbar 
\chi}{4}(\sigma_{1x}\sigma_{2x}+\sigma_{2x}\sigma_{3x}+\sigma_{3x}\sigma_{4x}+\sigma_{4x}\sigma_{1x})$ and ${\bf 
A}=\frac{{\bf H}}{\hbar \chi}$. 
The basis states $|j,m\rangle$ where $j=2$ and $m=-2,\,-1,\,0,\,1,\,2$ of the symmetric subspace are given by 
\begin{eqnarray}
\label{eq:jmstates}
\vert\phi_1\rangle&=&|2,2\rangle=|0000\rangle \nonumber \\ 
\vert\phi_2\rangle&=&|2,1\rangle=\frac{1}{2}\left[|0001\rangle+|0010\rangle+|0100\rangle+|1000\rangle\right] 
\nonumber \\
\vert\phi_3\rangle&=&|2,0\rangle=\frac{1}{2}\left[|0011\rangle+|1100\rangle+|0101\rangle+|1010\rangle+|0110
\rangle+|1001
\rangle\right] \\
\vert\phi_4\rangle&=&|2,-1\rangle=\frac{1}{2}\left[|1110\rangle+|1101\rangle+|1011\rangle+|0111\rangle\right] 
\nonumber \\
\vert\phi_5\rangle&=&|2,-2\rangle=|1111\rangle  \nonumber 
\end{eqnarray}

The time evolved states $\vert\phi'_\alpha\rangle={\bf U}\vert\phi_\alpha\rangle$ ($\alpha=1,\,2,\,3,\,4,\,5$) 
remain in the symmetric subspace iff $\vert\phi'_\alpha\rangle$ is expressible as 
$\vert\phi'_\alpha\rangle=\sum_{m} c_{m}|2,m\rangle$ such that $\sum_{m}|c_m|^2=1$. We have evaluated the set of 
coefficients $c_m=\langle\psi'_\alpha|2,m\rangle, \ m=-2,\,-1,\,0,\,1,\,2$ for each $\alpha$, 
($\alpha=1,\,2,\,3,\,4,\,5$) and these coefficients  are given explicitly in Table~II.

It is readily seen from Table~II that though all the states $\vert\phi_\alpha\rangle$ ($\alpha=1$ to $5$) are 
initially symmetric (at time t=0, $\sum_m c_m^2=1$ for all $\alpha$), their time-evolved counterparts 
$\vert\phi'_\alpha\rangle$ are not restricted to the symmetric subspace. After time evolution, 
$\vert\phi_2\rangle=|2,1\rangle$ and $\vert\phi_4\rangle=|2,-1\rangle$ the so-called W states, are the only two 
that remain in the symmetric subspace and hence are symmetric under interchange of particles. We thus conclude 
that not all four qubit symmetric states retain their exchange symmetry after Ising chain interaction. Only a subclass of 
symmetric states, of the form $a|2,1\rangle+b|2,-1\rangle$ where $a$, $b$ are any two complex numbers can retain 
their exchange symmetry under Ising chain evolution. 

It is important to notice here that though the four qubit W states retain their symmetry under Ising chain evolution, $N$ qubit W states ($N\geq 5$) do not possess this property of symmetry retention.
In fact, all the $N$ qubit symmetric states ($N\geq 5$), the states belonging to the $N+1$ dimensional symmetric subspace, lose their exchange symmetry on interaction with Ising chain with nearest neighbor interactions. The loss of exchange symmetry in $N$-qubit Dicke states ($N$ greater than 5) can be checked by inspection as is done for the 2 qubit case. The loss of symmetry is evident in just the action of the $N$-qubit Ising chain Hamiltonian on the corresponding Dicke 
states. Repeated applications of the Hamiltonian does not initiate  symmetry any further and the action of unitary time 
evolution operator corresponding to $N$-qubit Ising spin chain results in non-symmetric states. 

Though the cause of loss of exchange symmetry in initially symmetric Dicke states on time evolution with Ising chain interaction is not apparent, it may be of interest to notice that in all cases of permutation symmetry retention, ${\bf J}^2$, the square of the collective angular momentum operator commutes with the Hamiltonian. Going by the same lines one can see that all the $N$ qubit symmetric states ($N\geq 5$) lose their exchange symmetry on time evolution with anisotropic Heisenberg Hamiltonian, anisotropy being in either $x$ or $y$ direction\footnote{Anisotropy along $z$ direction does not spoil the symmetry of a symmetric $N$ qubit state because all the states spanning $N+1$ dimensional symmetric subspace are common eigenstates of ${\bf J}_z$ and ${\bf J}^2$ }. 

\section{Conclusion}
In this article we have shown that the symmetric $N$-qubit states  ($N\geq 5$) lose their exchange 
symmetry after interaction with a spin chain modelled by 1D Ising Hamiltonian with nearest neighbor 
interaction. Specifically we have shown that 2 and 3 qubit symmetric states retain their exchange symmetry under Ising chain evolution but all 4 qubit symmetric states, except the states of W type, lose their symmetry under the same interaction.  We emphasize here that permutation symmetry aspects are important either in studying the collective 
behaviour such as spin-squeezing or in relating the spin-squeezing behaviour of a symmetric N-qubit state with 
its pairwise entanglement properties. However, exchange symmetry properties are assumed to hold good under 1D Ising 
chain, when such studies are reported in Ref.~\cite{asmolmer,ssI1} and \cite{ssI2}. Through this work, we hope to initiate proper clarifications on the retention of exchange symmetry in future investigations on spin squeezed states.   

\begin{center}{\bf Acknowledgements}\end{center} 
We are grateful to the insightful suggestions by Prof. A. K. Rajagopal

\pagebreak
\begin{center}
\begin{table}
\label{table}
\caption{}
\vskip 0.1in
\begin{tabular}{|c|c|c|c|c|c|}
\hline
& & & & &  \\
State & $c_1=\langle \psi'_\alpha|3/2,3/2\rangle$ & $c_2=\langle \psi'_\alpha|3/2,1/2\rangle$ & $c_3=\langle 
\psi'_\alpha|3/2,-1/2\rangle$ & $c_4=\langle \psi'_\alpha|3/2,-3/2\rangle$ & $\sum_m c_m^2$ \\ 
& & & & &  \\ 
\hline
& & & & &  \\
$\vert\psi_1\rangle$ & $\cos^3 \frac{\chi t}{4} -i \sin^3 \frac{\chi t}{4}$  & 0 & $\frac{i\sqrt{3}}{2}e^{-i\chi 
t/4} \sin \frac{\chi t}{2}$ & 0 &  1 \\
& & & & &  \\ 
\hline
& & & & &  \\
$\vert\psi_2\rangle$ & 0 & $\frac{e^{3i\chi t/4}}{4}(3+e^{-i\chi t})$ &0 &$\frac{i\sqrt{3}}{2}e^{-i\chi t/4} 
\sin \frac{\chi t}{2}$ & 1 \\
\hline  
& & & & &  \\
$\vert\psi_3\rangle$ &$\frac{i\sqrt{3}}{2}e^{-i\chi t/4} \sin \frac{\chi t}{2}$ & 0 & $\frac{e^{3i\chi 
t/4}}{4}(3+e^{-i\chi t})$  & 0 & 1 \\ 
& & & & &  \\ 
\hline
& & & & &  \\
$\vert\psi_4\rangle$ & 0 & $\frac{i\sqrt{3}}{2}e^{-i\chi t/4} \sin \frac{\chi t}{2}$& 0 & $\cos^3 \frac{\chi 
t}{4} -i \sin^3 \frac{\chi t}{4}$ & 1\\
& & & & &  \\
\hline  
\end{tabular}
\end{table}
\end{center}
\begin{table}
\label{table}
\caption{}
\vskip 0.1in
\begin{tabular}{|c|c|c|c|c|c|c|}
\hline
& & & & & & \\
State & $c_1=\langle \phi'_\alpha|2,2\rangle$ & $c_2=\langle \phi'_\alpha|2,1\rangle$ & $c_3=\langle 
\phi'_\alpha|2,0\rangle$ & $c_4=\langle \phi'_\alpha|2,-1\rangle$ & $c_5=\langle \phi'_\alpha|2,-2\rangle$ & 
$\sum_m c_m^2$ \\ 
& & & & & & \\ 
\hline
& & & & & & \\
$\vert\phi_1\rangle$ & $\frac{1}{4}(3+\cos \chi \,t)$ & 0 & $\frac{-1 + \cos \chi\,t - 2i \sin \chi\,t}{2 
\sqrt{6}}$ & 0 & $\frac{1}{4}(-1+\cos \chi \,t)$ & $\frac{1}{6}(5+\cos \chi\,t)$ \\
& & & & & & \\ 
\hline
& & & & & & \\
$\vert\phi_2\rangle$ & 0 & $\frac{1 + \cos \chi\,t + i \sin \chi\,t}{2}$ & 0 & $\frac{-1 + \cos \chi\,t + i \sin 
\chi\,t}{2}$ & 0 & 1 \\
& & & & & & \\
\hline  
& & & & & & \\
$\vert\phi_3\rangle$ & $\frac{-1+\cos \chi \,t-2i\sin \chi \, t}{2\sqrt{6}}$ & 0 & $\frac{1 + 5\cos \chi\,t - 4i 
\sin \chi\,t}{\sqrt{6}}$ & 0 & $\frac{-1 + \cos \chi\,t - 2i \sin \chi\,t}{2\sqrt{6}}$ & $\frac{1}{9}(8+\cos 
\chi\,t)$ \\ 
& & & & & & \\ 
\hline
& & & & & & \\
$\vert\phi_4\rangle$ & 0 & $\frac{-1 + \cos \chi\,t + i \sin \chi\,t}{2}$ & 0 & $\frac{1 + \cos \chi\,t +i \sin 
\chi\,t}{2}$ & 0 & 1 \\
& & & & & & \\
\hline  
& & & & & & \\
$\vert\phi_5\rangle$ & $\frac{1}{4}(-1+\cos \chi \,t)$ & 0 & $\frac{-1 + \cos \chi\,t - 2i \sin \chi\,t}{2 
\sqrt{6}}$ & 0 & $\frac{1}{4}(3+\cos \chi \,t)$ & $\frac{1}{6}(5+\cos \chi\,t)$ \\ 
\hline
\end{tabular} 
\end{table}

\end{document}